# Naturally-Degradable Photonic Devices with Transient Function by Heterostructured Waxy-Sublimating and Water-Soluble Materials

*Andrea Camposeo[1,2], Francesca D'Elia[2], Alberto Portone[1,2], Francesca Matino[1,2], Matteo Archimi[1,3], Silvia Conti[4], Gianluca Fiori[4], Dario Pisignano[1,3], Luana Persano[1,2]*

[1]NEST, Istituto Nanoscienze-CNR, Piazza S. Silvestro 12, I-56127 Pisa, Italy

[2]NEST, Scuola Normale Superiore, Piazza S. Silvestro 12, I-56127 Pisa, Italy

[3]Dipartimento di Fisica, Università di Pisa, Largo B. Pontecorvo 3, I-56127 Pisa, Italy

[4]Dipartimento di Ingegneria dell'Informazione, Università di Pisa, Via Caruso 16, I- 56122 Pisa, Italy

Devices that undergo an univocally-evolving designed function, possibly including a self-elimination process to harmless end products through mechanical fragmentation and dissolution in water or air, are collectively referred as physically-transient electronics and photonics.[1,2] The interest in such technology is rapidly emerging, and various forms of transient electronics have been recently traced,[1] including dissolvable devices that deploy room-temperature liquid metals with high recycling efficiency.[3,4] Indeed, the continuous increase of electronic waste, that will be further enhanced by the daunting amount of sensors and devices to be used for the forthcoming Internet of Things,[5] is motivating significant concern. For these reasons, technologies are highly desired, which might produce naturally-degradable devices, that once ended their function do not lead to release of heavy metals or toxic compounds in the environment. Such feature is also of interest for other applications where conventional electronics and photonics are not well-suited such as biodegradable, temporary implants,[6] secure data-storage systems,[7] and unrecoverable remote controls.[8] For instance, physically-transient light emitting components provide an interesting option in various domains, such as biophotonics,[9] implantable analytic chips,[10] wireless optogenetics[11] and full-field imaging.[12] Examples of transient photonic systems include nanoimprinted distributed feedback lasers based on DNA[2] or silk gratings obtained by using fused silica,[13] and fluorescent chemosensors for acid vapor detection.[14] Furthermore, 'dry' forms of transient electronics have been recently developed for transistors or solar cells realized by transfer printing onto materials that sublime.[15] In fact, ambient sublimation of device-carrying substrates leads to mechanically disintegrate the remaining materials into microscopic fragments (due to loss of support and/or inter-layer adhesion). Dissolution in water could then be deployed to finally eliminate components at the end of their life cycle.

Here, we present the first example of combined dry-wet transient devices in photonics, based on water-soluble compounds layered onto sublimating substrates. Cyclododecane (CDD), a wax-like non polar cyclic hydrocarbon solid, widely used as a temporary consolidant, adhesive, or barrier during archeological recovery,[16] serves as substrate for an optically-active water-dissolvable polymer bi-layer encompassing a laser dye, as schematized in Figure 1a. The relevance of this method is twofold, involving both fundamental and practical findings. On one side the compatibility of sublimating substrates with organic photonics is here assessed, that goes beyond previous results on silicon-based electronics,[15] tackling potential issues related to deterioration of light-emitting molecules interacting with CDD or excess light-scattering. In addition, heterostructures built on the sublimating substrates are proposed, including both organic lasers and optical labels (QR-codes) with transient behavior and stably encoded information, showing two different application fields for naturally-degradable, all organic photonic devices.

As first step of the device fabrication process, grains of CDD paraffin are melted at 140°C and then rapidly cooled in a refrigerator to form substrates with size $2 \times 2$ cm$^2$ and thickness in the range 1-3 mm (Figure 1b). Silicon molds are used to generate substrates which can have varied shapes, different dye-doping (Figure 1b-d), and smooth surfaces (root mean square roughness, $R_q$= 70 nm). Alternative templates such as glass Petri dishes, and other methods explored such as applying mechanical forces to solid grains, lead to samples with higher roughness ($R_q$=25 μm and 770 nm, respectively, Figure S1).

The obtained CDD substrates are soft, and can be cut into pieces of the desired size and shape by the use of a scalpel, as shown in Figure 1e,f. Mechanical analysis in compression mode, performed by thrusting downward a flat disk onto the substrate at a rate of 0.5 mm/min at room temperature, highlights a roughly linear force-deformation response i.e. an elastic modulus of (14±1) MPa up to about 30 N (~1-1.5% compression), followed by irreversible deformation indicative of the material transport and ductile behaviour (Figure S2). At ambient temperature under a fume hood with an air velocity of 1.9 m s$^{-1}$, and relative humidity of about 50%, a substrate (thickness 1-2 mm) completely disappears in a time of 7-9 days (Figure 1g-i), corresponding to an area reduction rate of about (0.6± 0.1) mm$^2$ hr$^{-1}$ and to a thickness reduction rate of about (12±1) μm hr$^{-1}$ (Figure 1j,k and Figure S3). These rates can be controlled to high extent, thus providing routes to realize components with programmable self-destruction, depending on the environment conditions where devices are put to use. For instance, sublimation rates can be enhanced by increasing the substrate temperature (Figure 1l,m; area reduction rate about 3.5 mm$^2$ hr$^{-1}$ and thickness reduction rate about 70 μm hr$^{-1}$), or considerably decreased by addition of nanoparticles.[15] In addition, the fabrication method also influences the subsequent rate of sublimation. Sublimation rates for compact and dense films made by applying a pressure (e.g. 40-60 kPa) onto CDD grains are about seven times lower than for the samples cooled from melt.

To analyze the sublimation process at microscopic scale and capture the evolution of the CDD material in real time, free-standing substrates are investigated by photon backscattering (PB) through confocal microscopy at ambient conditions. The samples are illuminated by a 488 nm cw laser, and videos of the backscattered signals are captured for about a few hours. Exemplary frames recorded at 0, 60 and 120 minutes are displayed in Figure 2a-c, providing the surface-reflected optical signal on a lengthscale of hundreds of microns. The backscattered intensity is found to decay with a mean lifetime of about 90 minutes (Figure 2d) and can be correlated to surface diffusion phenomena and sublimation that while proceeding moves the CDD material out of the detection focal plane. Indeed, the corresponding rate (order of 10 μm hr$^{-1}$) can be estimated by taking into account the focal depth of the confocal microscope[17] (see Experimental) and assuming that the sublimation from the region of interest is completed when the backscattered signal zeros. Cross-polarized optical microscopy is also used to investigate the CDD film appearance, since paraffin waxes are typically composed of anisotropic crystals with size and shape strongly depending on processing history and in particular on the cooling rate from melt. For instance, long needle-like crystals are favored in slow cooling conditions while smaller round shaped crystals are favored when the material is quickly cooled from melt by using liquid nitrogen or by fast refrigeration.[18] Here the morphology, observed for 8 hours in ambient conditions, shows mainly rounded material domains for which an average area of 2-3×10$^3$ μm$^2$ is imaged within the microscope depth of focus (Figure 2e-g). Though in absence of significant sample heating (Figure S4), this average value slowly varies over time (by ~10$^2$ μm$^2$ hr$^{-1}$), related to free volume redistribution during sublimation.

Light-emitting heterostructures that can operate as optical amplifiers are built by drop-casting bilayers made of poly(vinyl alcohol) (PVA) and polyvinylpyrrolidone (PVP) doped with the red-emitting laser dye [2-[2-[4-(dimethylamino)phenyl]ethenyl]-6-methyl-4H- pyran-4-ylidene]-propanedinitrile (DCM), onto the CDD substrate, according to the scheme in Figure 1a. The device cross-section is shown in the white light and fluorescence micrographs in Figure 3a,b, where a sample with the intermediate PVA intentionally doped with a blue-emitting molecule (stilbene 420) is used to better highlight the neat interfaces between the different layers. The absorption and photoluminescence (PL) spectra of the active film deposited on a reference quartz substrate are displayed in Figure S5. During device fabrication, orthogonal solvents allow the transient heterostructures to be built by sequential casting. Specifically, water is used to dissolve PVA, thus enabling the formation of a continuous planarizing coating on top of the CDD. The PVA-CDD interface is found to be stable, and delamination is only occasionally observed from sample corners during heterostructure sample cutting (surface area ~1 cm$^2$), which could be related to undesired air pockets trapped at the interface formed during polymer casting. Furthermore, since CDD is soluble in polar solvents, the direct deposition of DCM:PVP from a chloroform solution is unviable onto the sublimating material, since this would lead to permeation of both the polymer and the dye into the substrate, as well as to a significant alteration of the dye spectral properties (Figure S6). Obtaining interfaces of high quality is clearly important for promoting waveguiding[19,20] of the photoexcited light emitted by the heterostructure, occurring mostly in the dye-doped layer (see Experimental). In addition, these heterostructures are compatible with encapsulation methods with optical adhesives, that can be used to strongly enhance the device photo-stability.[21]

The PL spectra, obtained from the device (Figure 3c) upon pumping with ~10 ns pulses at 355 nm and with a stripe excitation geometry (Experimental), have a linewidth (full width at half maximum, FWHM) down to 25 nm, and a threshold at 0.2 mJ cm$^{-2}$ for amplified spontaneous emission (ASE) as displayed in the light-light (*L-L*) plot in Figure 3d. The spectra do not exhibit intrinsic polarization (Figure S7), which is indicative of an isotropic distribution obtained for the dye chromophores upon casting. In Figure 3e, we display the output beam divergence, namely the dependence of the intensity of the emitted light on the observation angle, that is measured by the experimental geometry shown in Figure S8. The spatial profile for ASE (Fig. 3e) evidences the coexistence of different light modes which are out-coupled, which is typical of asymmetric emissive layers embedding organic layers with optical gain.[22] Furthermore, the device photostability in air is studied by recording spectra of the emitted light as a function of the number of excitation pulses (i.e., time), at a constant pump fluence of 0.3 mJ cm$^{-2}$ (Figure 3f). The operational lifetime of the device (given by the number of excitation pulses at which the ASE intensity decays by $1/e$) is so estimated to correspond to about $1.5 \times 10^3$ pulses, with a FWHM increasing by a factor 1.5 in a related way (Figure 3g). This performance is fully in line with the photostability measured for organic lasers using well-established architectures, including distributed feedback resonators.[23,24]

The possible applications of these devices before controlled disintegration are numerous. They can be used as high-quality, bright illumination sources for full-field imaging, since unconventional laser architectures with low spatial coherence allow for producing images without speckle artefacts.[25] In previous works, random lasers made of colloidal solutions or perovskite films have been proposed as tool for speckle-free imaging.[12,26,27] In order to analyze this capability for ASE transient devices, we use them to illuminate a 1951 U.S. Air Force (AF) resolution test chart, and compare so-obtained images with those produced by a narrowband laser (Nd:YAG). The experimental set-up is schematized in Figure S9. The illuminated images of the test patterns are shown in Figure 3h,i. While textured artefacts due to speckle interference are found in the image collected from by Nd:YAG laser illumination (Figure 3h), the ASE from the transient device, providing an optical signal with low spatial coherence, allow clean imaged patterns to be achieved, free from speckle effects (Figure 3i). A quantitative comparison can be carried out by estimating the contrast to noise ratio, CNR=$(\langle I_f \rangle - \langle I_b \rangle)/[(\sigma_f + \sigma_b)/2]$, where $\langle I_f \rangle$ is the average intensity of a reference bar in the test chart, $\langle I_b \rangle$ is the average intensity of the image background, and $\sigma_{f,b}$ indicate the corresponding standard deviations of the pixel intensity. By using the transient device as light source, CNR values are up to four times better than by using a conventional laser and comparable to those obtained by colloidal or perovskite lasers (Figure 3j).[26,27]

Upon sublimation the CDD material loss can deeply modify the device morphology. To capture this behavior we perform accelerated sublimation tests by placing the ASE transient device on a hot plate at 55°C and imaging it by both an infrared thermographic camera (Figure 4a) and by optical photographs (Figure 4b). The temperature is chosen to avoid local melting (occurring above 60°C) of the surface in contact with the hot plate. A temperature gradient of about 3 °C mm$^{-1}$ is measured across the CDD substrate, keeping the active layer at a temperature significantly lower than the heated parts. In addition, it is found that sublimation proceeds faster at the edges of the overall heterostructure, which is gradually led to take a mushroom-like shape with the light-emitting region remaining largely unaltered at the top of the device. Bending of the PVA/DCM:PVP bilayer could occur as a consequence of the substrate sublimation and the consequent strain energy release in the two polymers. The amount of bending would depend on the thickness ratio ($R$) of the two layers (DCM:PVP thickness/PVA thickness), being less pronounced upon decreasing $R$ below unity.[28] In Figure 4c, we show how this aspect can be engineered to keep the light-emitting structure mechanically stable and unbent, even upon complete CDD sublimation, by properly choosing the thickness ratio ($R$=0.4). Such $R$ value engineered for PVA/DCM:PVP can depend on the environmental relative humidity, due to differential water vapor absorption and polymer swelling for the two involved layers. Keeping the optically-relevant structure mechanically stable is highly desired in order to use the photonic structure with unchanged overall geometry and output beam directionality along the component lifetime. For instance, this property is highly relevant to realize intelligent, optical labels that are naturally degraded after read while keeping stably readable the information encoded in them during use. We provide a demonstration of this behaviour by realizing QR-code patterns through inkjet printing of poly(3,4-ethylenedioxythiophene):polystyrene sulfonate (PEDOT:PSS) onto CDD/PVA/PVP heterostructures (Figure 5). To this purpose, a Dimatix Materials Printer 2800 (Fujifilm) is employed, developing a direct, drop-on-demand printing process entirely at room temperature in air. A 1 pL cartridge is used with a drop spacing of 20 μm, while printing six overlapped PEDOT:PSS layers in order to improve the optical contrast in the labels. The method is mask-less, with reduced material waste and good lateral resolution (tens of micrometers), well-matching the requirements of optical labels. The deposition of the PEDOT:PSS layers locally increases the optical density in the visible range (400-700 nm) by about a factor 3 (from 0.4 to 1.3, corresponding to a decrease of the intensity of light transmitted by CDD/PVA/PVP/PEDOT:PSS by an order of magnitude with respect to pristine CDD/PVA/PVP). This allows the patterns to be detectable by a smartphone-embedded camera (Figure 5a,d,g,j). During the sublimation of the CDD substrate, the devices do not significantly bend, making the information encoded in them easily readable by the smartphone scanner (Figure 5c, f, i, l, where the words 'Crypted Message' correspond to the reading of the QR-code patterns at each time step).

Finally, after device use and complete self-destruction of the CDD layer, the QR-codes are disintegrated and the remaining polymer layers straightforward dissolved in water (Figure 5m-p and Figure S10). While ultimately dissolving the exhausted components, water is not functional to the device operation, thus avoiding uncontrolled flows or undesired variability related to dye quenching or dissolution rates. In this respect, dry transient technologies provide additional self-eliminating capability, that complements those of widely-explored wet systems, and is critically useful in environments where little water or moisture are present.

To summarize, our work introduces new physically-transient photonic devices with programmable disintegration. Compared to existing transient architectures involving semiconductors for electronics, bioelectronics, and photovoltaics,[15,29,30] these systems are based on fully organic components, with extremely low cost associated to the related processing technologies and absence of dissolved metals. The combined use of wet and dry transient materials might lead to systems with further enhanced and versatile transient functions, and to new fabrication routes for organic lasers[31] and encoded labels, among others, with potential application in environmental sensing, storage conditions monitoring, and photonic chips. For instance, the here analysed time scales (a few days) can be useful for making transient fluorescence excitation sources in disposable lab-on-chip platforms,[32,33] where the transient devices can be directly coupled with integrated waveguides and interrogation regions. Importantly, the method is also of interest in terms of environmental safety, suppressing costs and hazards associated with discarded devices.

*Experimental Section*

*Transient Devices.* CDD was purchased from Kremer Pigmente GmbH & Co. KG (Aichstetten). PVA and PVP were purchased from Sigma Aldrich. DCM was purchased from Exciton Inc.. PEDOT:PSS (Clevios[TM] PH 1000 purchased from Heraus) was used as antistatic waterbone dispersion based on submicrometer gel particles. Upon CDD melting, substrates were rapidly refrigerated from the liquid phase to obtain a dense amorphous phase. The surface roughness of CDD obtained by using different molds was determining by inspecting sample profile by a stylus profilometer (Bruker Dektak XT). Mechanical measurements in compression mode were performed by a mechanical tester (INSTRON 5500R) with stainless still parallel plates (diameter 5 cm) and equipped with MERLIN software (INSTRON; Norwood, MA, USA). A bilayer of polymer PVA/DCM:PVP (1% wt:wt) of DCM was deposited by casting directly on top of the CDD substrate. The refractive indexes of involved layers (~1.48 for CDD and PVA and ~1.53 for PVP) promote waveguiding only in the dye-doped film. The thickness of each deposited polymer layer, in the range 45-90 μm (PVA) and ~30 μm (DCM:PVP), was measured by observing cross-sections by an optical stereo-microscope in bright and dark field (Leica MZ16 FA). The method was also supported by looking at identical samples upon fluorescence excitation, by doping PVA with the blue-emitting dye stilbene 420 (Exciton inc.). Transient QR-codes were realized by direct drop-on-demand, inkjet printing of PEDOT:PSS. A Dimatix Materials Printer 2800 (Fujifilm) was used under ambient conditions, with a 16-nozzle cartridge by 1pL single-drop volume, a single nozzle to improve accuracy, and drop spacing of 20 μm. Six overlapping layers of PEDOT:PSS were printed, with a 60 s pause between two consecutive printing steps to allow for effective drying, thus guaranteeing the deposition of a uniform, dark film in the label features.

*Microscopy.* An inverted microscope, Eclipse Ti (Nikon), equipped with a confocal A1R-MP system (Nikon) was used for the microscopic optical measurements. For the characterization of the light backscattered by the CDD material, a polarized Ar ion laser (wavelength, $\lambda = 488$ nm) was used as the light source, while the backscattered signal was collected by a 10× objective (numerical aperture, N.A.=0.25) and measured by a photomultiplier. At the beginning of the measurement, the objective is positioned in order to have the focal plane at the surface of the sample (air/sample interface), and this objective position is fixed throughout the overall measurement. Upon sublimation from the sample, the light intensity that is backscattered by the air/sample interface and measured by the detector of the confocal microscope is to decrease because of the increasing spatial gap between the air/sample interface and the fixed focal plane. The sublimated thickness was estimated by the focal depth of the confocal systems: $\Delta z = 2\lambda / (N.A.)^2$.[17] The bright field imaging time-lapse was collected by a DS-Ri1 color charged-coupled-device (CCD) camera (Nikon) in cross-polarization mode. Analysis was performed by the ImageJ software. QR-codes were imaged by using either a smartphone camera or a CCD camera (Leica) coupled to a long working-distance optical system (MVL7000, Thorlabs).

*Thermographic imaging.* Thermal imaging of the devices was carried out by using an infrared camera (FLIR, A655sc). Before each thermographic measurement[34], the camera was calibrated following the manufacturer guidelines.

*Spectroscopy.* A JASCO UV-VIS spectrophotometer V-550 was used to measure absorption spectra of the active bilayers, while PL spectra were acquired by a Cary Eclipse Fluorescence Spectrophotometer (Varian). The transient photonic devices were excited by the third harmonic ($\lambda=355$ nm) of a Nd:YAG laser (Quanta-Ray INDI, Spectra-Physics) with 10 ns pulse duration (FWHM) and 10 Hz repetition rate. The excitation beam was shaped into a stripe with size 5×0.5 mm². The intensity and spectral properties of the light emitted from the sample edge were characterized by a fiber-coupled monochromator (iHR320, JobinYvon) equipped with a Peltier-cooled CCD detector (Symphony, Horiba) and a 600 grooves×mm$^{-1}$ grating. A rotating polarizer positioned along the path of the collection of the light emitted by the sample was used for analyzing the polarization state of the emission. The divergence of the light beam emitted by the device was determined by using a mask with a slit aperture (width 1 mm and height 10 mm) positioned perpendicularly to the emission direction (Figure S8), and measuring the intensity of the light transmitted through the slit as a function of the transversal position of the mask. A negative 1951 US Air Force resolution test chart (Thorlabs) was used for imaging experiments performed with the transient photonic device and with a laser light source (Figure S9). Images were collected with a 10× objective lens (M-10X, Newport, N.A.=0.25) and a CMOS camera (Thorlabs).

*Statistical Analysis.* The area and thickness of CDD substrates during sublimation (Figure 1 and Figure S3) were normalized at values measured at $t$=0. Sublimation rates at ambient temperature were averaged over 15 different samples. Experiments to measure the elastic modulus were repeated on three different samples realized with the same processing conditions. The lasing behavior was analyzed on five different devices. The average values of CNR (Figure 3j) was estimated on six pattern lines at each group size.

*Acknowledgements*

The research leading to these results has received funding from the European Research Council (ERC) under the European Union's Horizon 2020 research and innovation programme (grant agreements No. 682157, "*x*PRINT" and No. 770047, "PEP2D"), from the H2020 WASP project (contract no. 825213), and from the Italian Minister of University and Research PRIN 2017PHRM8X and PRIN 20173L7W8K projects. Sureeporn Uttiya is acknowledged for her support in light-emitting substrate fabrication with silicon molds. Alina Adamow is acknowledged for her assistance during absorption and PL measurements. Serena Danti and Luca Panariello at the Multifunctional Bio-Ecocompatible Materials Laboratory (MBEM), Dept. of Civil and Industrial

Engineering, University of Pisa) are acknowledged for support by mechanical measurements. L.P. acknowledges Dan Franklin and John A. Rogers for helpful suggestions and fruitful discussion.

## References

[1] J.-K. Chang, H. Fang, C. A. Bower, E. Song, X. Yu, J. A. Rogers, *Proc. Natl. Acad. Sci. USA* **2017**, *114*, E5522.

[2] A. Camposeo, P. Del Carro, L. Persano, K. Cyprych, A. Szukalski, L. Sznitko, J. Mysliwiec, D. Pisignano, *ACS Nano* **2014**, *8*, 10893.

[3] L. Teng, S. Ye, S. Handschuh-Wang, X. Zhou, T. Gan, X. Zhou, *Adv. Function. Mater.* **2019**, *29*, 1808739.

[4] R. Guo, X. Sun, B. Yuan, H. Wang, J. Liu, *Adv. Sci*. **2019**, *6*, 1901478.

[5] S. Fang, L. D. Xu, Y. Zhu, J. Ahati, H. Pei, J. Yan, Z. Liu*, IEEE Trans. Industr. Inform.* **2014**, *10*, 1596.

[6] D. Son, J. Lee, D. J. Lee, R. Ghaffari, S. Yun, S. J. Kim, J. E. Lee, H. R. Cho, S. Yoon, S. Yang, S. Lee, S. Qiao, D. Ling, S. Shin, J. K. Song, J. Kim, T. Kim, H. Lee, J. Kim, M. Soh, N. Lee, C. S. Hwang, S. Nam, N. Lu, T. Hyeon, S. H. Choi, D. H. Kim, *ACS Nano* **2015**, *9*, 5937.

[7] G. Lee, S.-K. Kang, S. M. Won, P. Gutruf, Y. R. Jeong, J. Koo, S.-S. Lee, J. A. Rogers, J. S. Ha, *Adv. Energy Mater* **2017**, *7*, 1700157.

[8] C. W. Park , S.-K. Kang, H. L. Hernandez, J. A. Kaitz, D. S. Wie, J. Shin, O. P. Lee, N. R. Sottos, J. S. Moore, J. A. Rogers, S. R. White, *Adv. Mater.* **2015**, *27*, 3783.

[9] E. A. Specht, E. Braselmann, A. E. Palmer, *Ann. Rev. Physiol*. **2017**, *79*, 93.

[10] K. Scholten, E. Meng, *Lab Chip* **2015**, *15*, 4256.

[11] G. Shin, A. M. Gomez, R. Al-Hasani, Y. R. Jeong, J. Kim, Z. Xie, A. Banks, S. M. Lee, S. Y. Han, C. J. Yoo, J.-L. Lee, S. H. Lee, J. Kurniawan, J. Tureb, Z. Guo, J. Yoon, S.-I. Park, S. Y. Bang, Y. Nam, M. C. Walicki, V. K. Samineni, A. D. Mickle, K. Lee, S. Y. Heo, J. G. McCall, T.

Pan, L. Wang, X. Feng, T.-i Kim, J. K. Kim, Y. Li, Y. Huang, R. W. Gereau IV, J. S. Ha, M. R. Bruchas, J. A. Rogers, *Neuron* **2017**, *93*, 509.

[12] B. Redding, M. A. Choma, H. Cao, *Nat. Photon.* **2012**, *6*, 355.

[13] H. Jung, K. Min, H. Jeon, S. Kim, *Adv. Opt. Mater.* **2016**, *4,* 1738.

[14] K. Min, S. Kim, C. G. Kim, S. Kim, *Sci. Rep.* **2017**, *7*, 5448.

[15] B. H. Kim, J.-H. Kim, L. Persano, S.-W. Hwang, S. Lee, J. Lee, Y. Yu, Y. Kang, S. M. Won, J. Koo, Y. K. Cho, G. Hur, A. Banks, J.-K. Song, P. Won, Y. M. Song, K.-I. Jang, D. Kang, C. H. Lee, D. Pisignano, J. A. Rogers, *Adv. Function. Mater.* **2017**, *27*, 1606008.

[16] S. Muñoz-Viñas, V. Vivancos-Ramón, P. Ruiz-Segura, *Restaurator* **2016**, *37*, 29.

[17] L. Novotny, B. Hecht, Principle of Nano-Optics, Cambridge University Press, Cambridge, UK 2006, p. 86.

[18] J. W. Bowen, T. Owen, J. B. Jackson, G. C. Walker, J. F. Roberts, D. M.-Levif, P. Lascourrèges, D. Giovannacci, V. Detalle, *IEEE Trans. THz Sci. Technol* **2005***, 5*, 1005.

[19] F. Hide, M. A. Diaz-Garcia, B. J. Schwartz, M. R. Andersson, Q. B. Pei, *Science* **1996**, *273*, 1835.

[20] A. K. Sheridan, G. A. Turnbull, A. N. Safonov, I. D. W. Samuel, *Phys. Rev. B* **2000**, *62*, 929.

[21] S. Richardson, O. P. M. Gaudin, G. A. Turnbull, and I. D. W. Samuel, *Appl. Phys. Lett.* **2007**, *91*, 261104.

[22] J. Peter, R. Prabhu, P. Radhakrishnan, C. P. G. Vallabhan, V. P. N. Nampoori, M. Kailasnath, *J. Appl. Phys.* **2015**, *117*, 015301.

[23] P. Del Carro, A. Camposeo, R. Stabile, E. Mele, L. Persano, R. Cingolani, D. Pisignano, *Appl. Phys. Lett*. **2006**, *89*, 201105.

[24] C. Vannahme, S. Klinkhammer, M. B. Christiansen, A. Kolew, A. Kristensen, U. Lemmer, T. Mappes, *Opt. Exp.* **2010**, *18*, 24881.

[25] Y.-J. Lee, T.-W. Yeh, Z.-P. Yang, Y.-C. Yao, C.-Y. Chang, M.-T. Tsai, J.-K. Sheu, *Nanoscale* **2019**, *11*, 3534.

[26] Y. Liu, W. Yang, S. Xiao, N. Zhang, Y. Fan, G. Qu, Q. Song, *ACS Nano* **2019**, *13*, 10653.

[27] Y.-C. Wang, H. Li, Y.-H. Hong, K.-B. Hong, F.-C. Chen, C.-H. Hsu, R.-K. Lee, C. Conti, T. S. Kao, T.-C. Lu, *ACS Nano* **2019**, *13*, 5421.

[28] A. I. Egunov, J. G. Korvink, V. A. Luchnikov, *Soft Matter* **2016**, *12*, 45.

[29] S.-W. Hwang, H. Tao, D.-H. Kim, H. Cheng, J.-K. Song, E. Rill, M. A. Brenckle, B. Panilaitis, S. M. Won, Y.-S. Kim, Y. M. Song, K. J. Yu, A. Ameen, R. Li, Y. Su, M. Yang, D. L. Kaplan, M. R. Zakin, M. J. Slepian, Y. Huang, F. G. Omenetto, J. A. Rogers, *Science* **2012**, *337*, 1640.

[30] M. Muskovich, C. J. Bettinger, *Adv. Healthc. Mater.* **2012**, *1*, 248.

[31] A. J. C. Kuehne, M. C. Gather, *Chem. Rev*. **2016**, *116*, 12823.

[32] A. Pais, A. Banerjee, D. Klotzkin, I. Papautsky, *Lab Chip* **2008**, *8*, 794.

[33] B. Ibarlucea, E. Fernandez-Rosas, J. Vila-Planas, S. Demming, C. Nogues, J. A. Plaza, S. Büttgenbach, A. Llobera, *Anal. Chem.* **2010**, *82*, 4246.

[34] R. Usamentiaga, P. Venegas, J. Guerediaga, L. Vega, J. Molleda, F. G. Bulnes, *Sensor* **2014**, *14*, 12305.

**Figures**

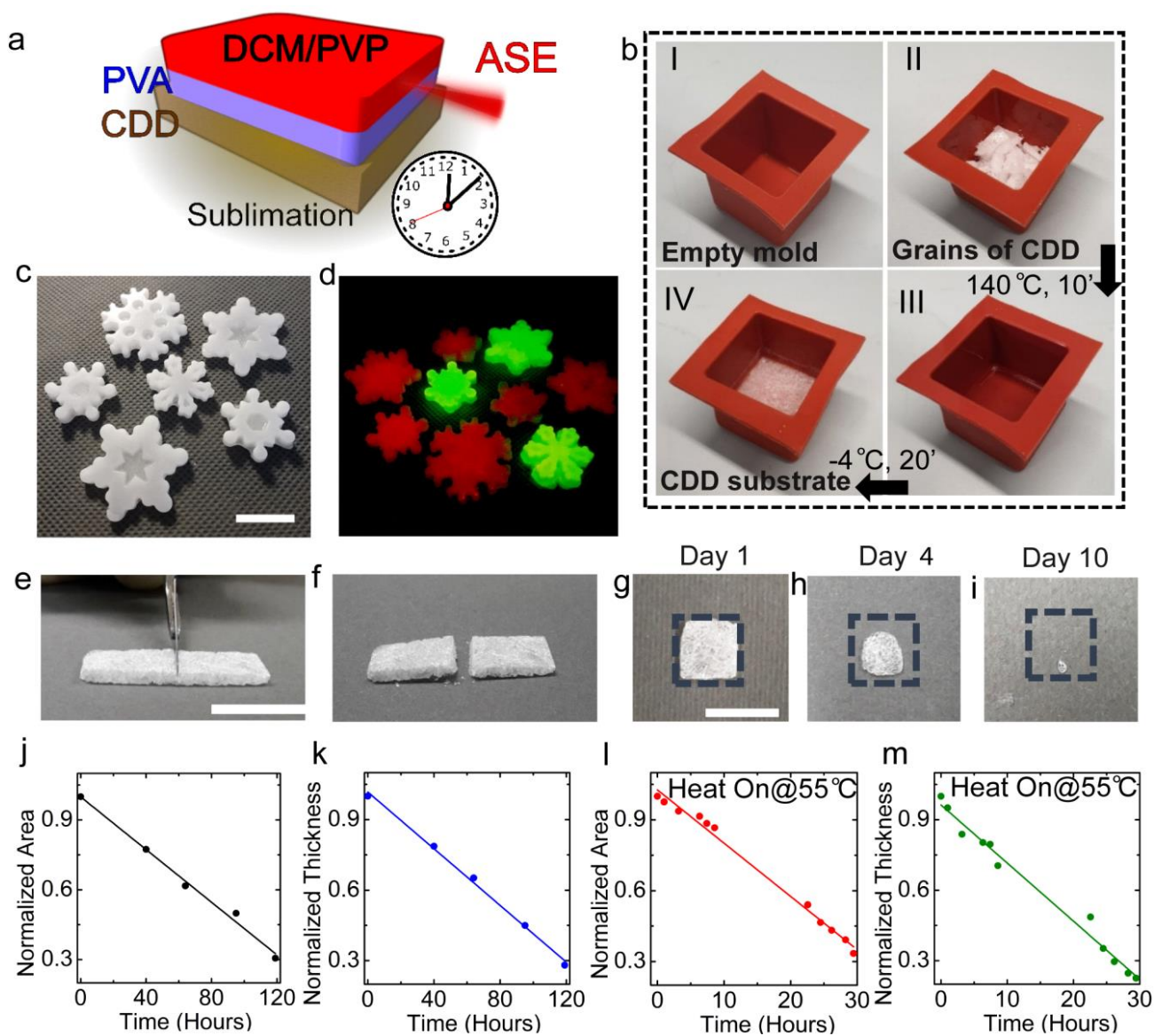

**Figure 1.** Preparation of CDD substrates with arbitrary shape, and their rate of sublimation. (a) Schematics of transient photonic heterostructures on a sublimating substrate. (b) Preparation of a CDD substrate in silicon molds, through steps I-IV. (c) Photographs of various substrate shapes obtained by the silicon molds. Scale bar: 3 cm. (d) CDD samples doped by dyes emitting different colours, i.e. rhodamine (red) and coumarine (green), imaged under UV illumination. (e,f) CDD parallelepipeds cut into into pieces of the desired size by using a scalpel. Scale bar in (e): 1 cm. (g-i) Photographs of substrates at different representative times. Scale bar: 1 cm. Area (j) and thickness (k) of CDD substrates during sublimation at room temperature, and at 55°C (l,m). Air velocity: 1.9 m s$^{-1}$.

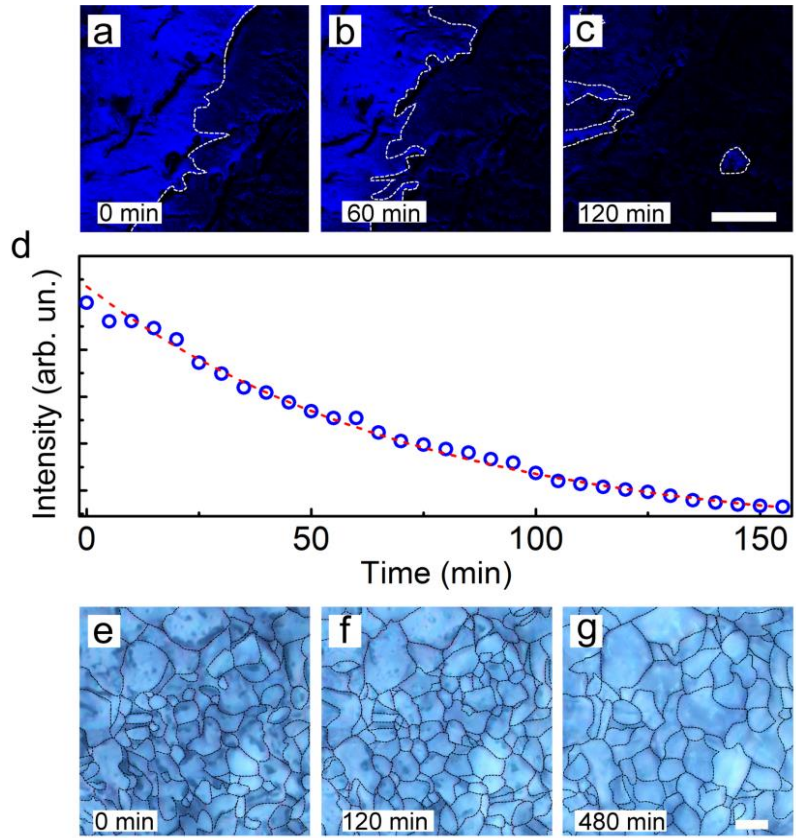

**Figure 2.** Microscopic images of a CDD region during sublimation. (a-c) PB confocal micrographs at different times. The white dashed lines highlight changes in the edges of the back-scattering material surface. Scale bar: 100 μm. (d) PB intensity *vs.* time, indicative of the sublimation behaviour. The red dashed line is a guide for the eyes. (e-g) Cross-polarized white light micrographs of CDD substrate at different times, indicative of dimensional changes in the material domains within the depth of focus. Scale bar: 100 μm.

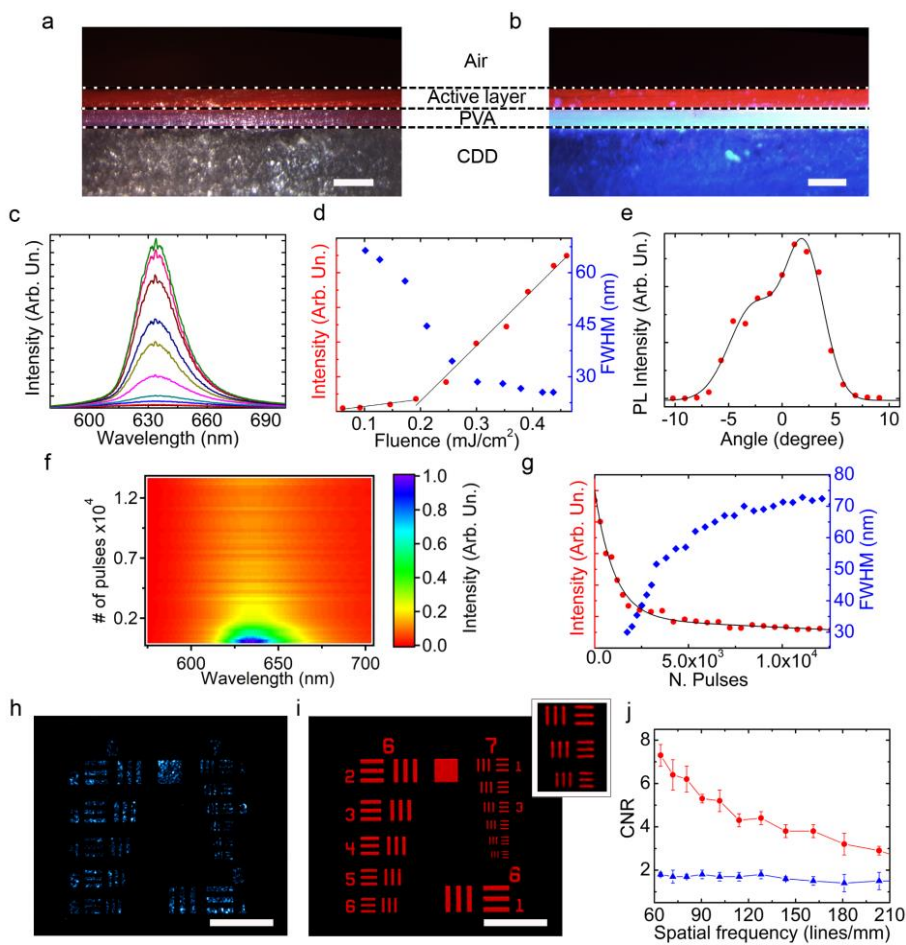

**Figure 3.** Transient photonic device. (a,b) Optical and fluorescence micrographs of the device cross section. The fluorescence image is obtained upon doping the PVA layer with stilbene (blue-emitting). Marker: 100 µm. (c) ASE spectra at different excitation fluences in the range: 0.06-0.5 mJ cm$^{-2}$. (d) Corresponding *L-L* plot (circles, left vertical scale), and dependence of the FWHM on the excitation fluence (diamonds, right scale). (e) Angular dependence of the emitted intensity (dots). The continuous line is a guide for the eye. (f) Dependence of the emission spectrum on number of pulses at 0.3 mJ cm$^{-2}$. (g) ASE intensity (red circle, left vertical scale) and FWHM (blue diamonds, right vertical scale) *vs.* the number of excitation pulses (repetition rate: 10 Hz) at 0.3 mJ cm$^{-2}$. The black line is a guide for the eye. (h,i). Images of a US Air Force resolution test chart taken by an Nd:YAG laser (h) and by the ASE from the transient device (i). Scale bar: 115 µm. Top right inset: magnification view of the test patterns in column 7, lines 4-6. (j) Comparison of CNRs for test patterns imaged by a narrow band laser (blue triangles) and by a transient photonic device (red dots). Average values estimated on 6 pattern lines at each group size.

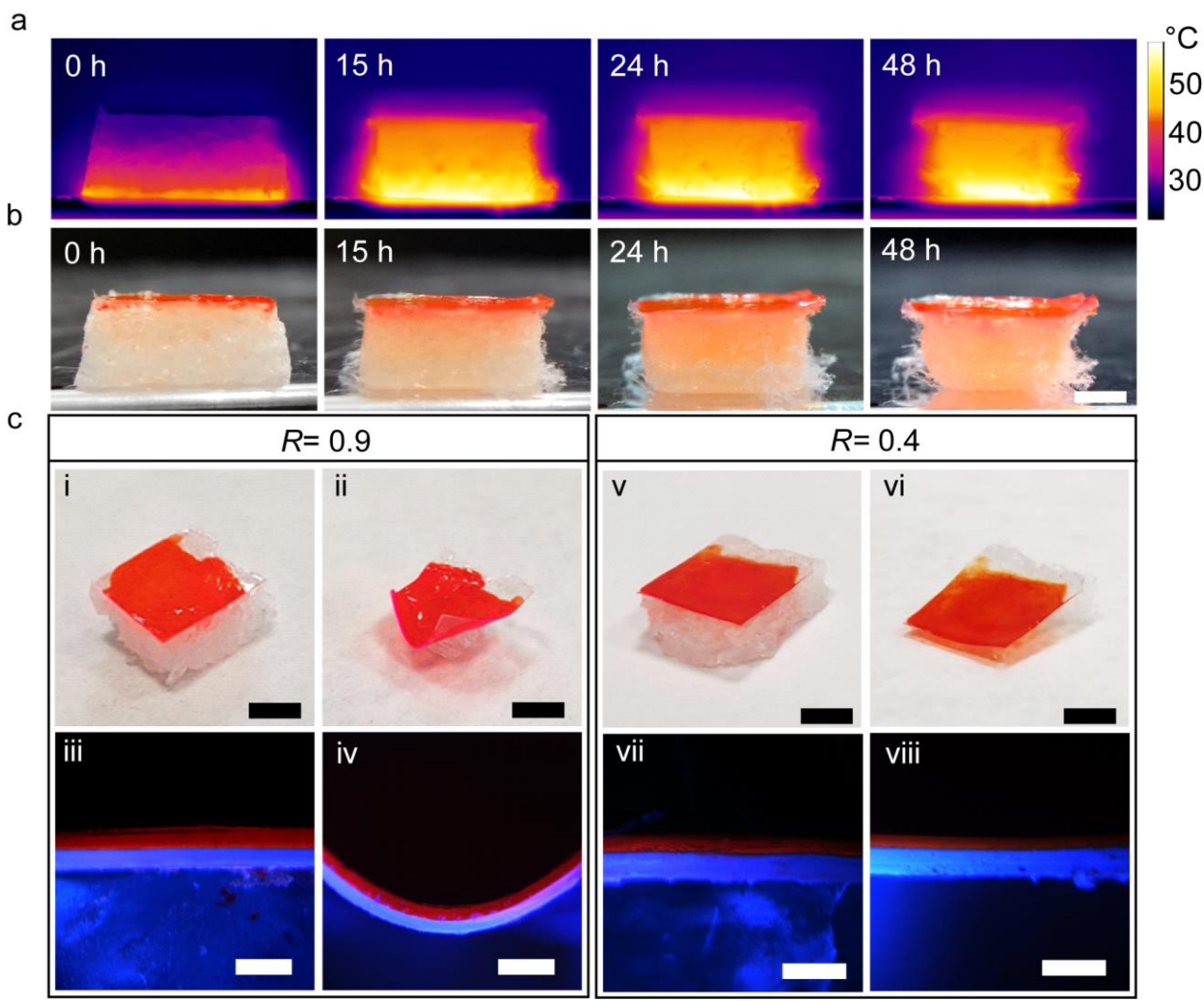

**Figure 4.** Transient device sublimation and dissolution. (a) Infrared images and corresponding photographs (b) of the ASE transient device while the substrate sublimates on a hot plate at 55°C. Four instants are captured during 48 hours. Scale bar: 2 mm. (c) Bending behaviour of transient devices, realized with different thickness ratio ($R$) of the two polymer layers, upon sublimation of the CDD substrate. Photographs (top line, scale bar: 0.5 cm) and fluorescence images of the cross-sections (bottom line, scale bar: 200 μm) of sample with $R$=0.9 (i-iv) and $R$=0.4 (v-viii). Samples are inspected before (i, iii, v, vii) and after (ii, iv, vi, viii) sublimation, respectively.

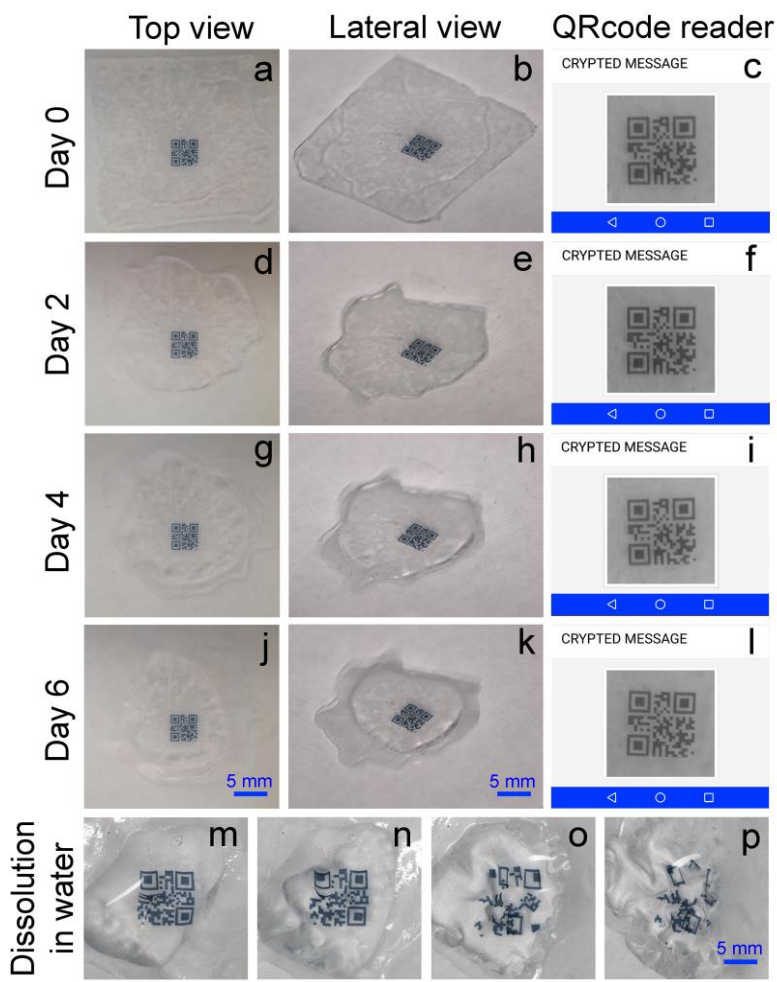

**Figure 5.** Top view (a, d, g, j) and lateral view (b, e, h, k) of a QR-code printed on CCD/PVA/PVP. The sample is kept in a fume hood with constant airflow (2 m/s) and monitored for one week, acquiring photographs at day 0 (a-c), 2 (d-f), 4 (g-i) and 6 (j-l). Top-view images are collected by a smartphone camera. Lateral-view images are collected by means of a CCD camera coupled to an optical system with long working distance. (c, f, i, l): Outputs of a QR-code reader after imaging the sample at day 0 (c), 2 (f), 4 (i) and 6 (l), respectively. The reader decodes correctly the label, each time reporting the text associated with it ('Crypted message'). (m-p): QR- code disintegration in water after the complete sublimation of the CDD substrate.

# Supporting Information

**Naturally-Degradable Photonic Devices with Transient Function by Heterostructured Waxy-Sublimating and Water-Soluble Materials**

*Andrea Camposeo[1,2], Francesca D'Elia[2], Alberto Portone[1,2], Francesca Matino[1,2], Matteo Archimi[1,3], Silvia Conti[4], Gianluca Fiori[4], Dario Pisignano[1,3], Luana Persano[1,2]*

[1]NEST, Istituto Nanoscienze-CNR, Piazza S. Silvestro 12, I-56127 Pisa, Italy

[2]NEST, Scuola Normale Superiore, Piazza S. Silvestro 12, I-56127 Pisa, Italy

[3]Dipartimento di Fisica, Università di Pisa, Largo B. Pontecorvo 3, I-56127 Pisa, Italy

[4]Dipartimento di Ingegneria dell'Informazione, Università di Pisa, Via Caruso 16, I- 56122 Pisa, Italy

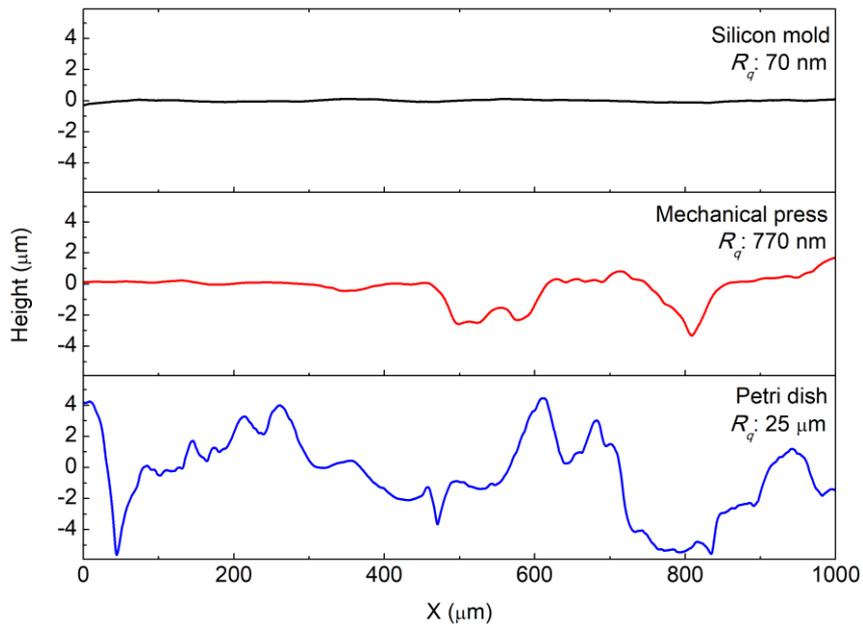

**Figure S1**. Height profiles of CDD surfaces upon molding by different techniques. From top to bottom: Heating-cooling cycles in pre-formed silicon molds, mechanical pressing of solid grains, heating-cooling in Petri dishes. $R_q$: calculated root mean square roughness.

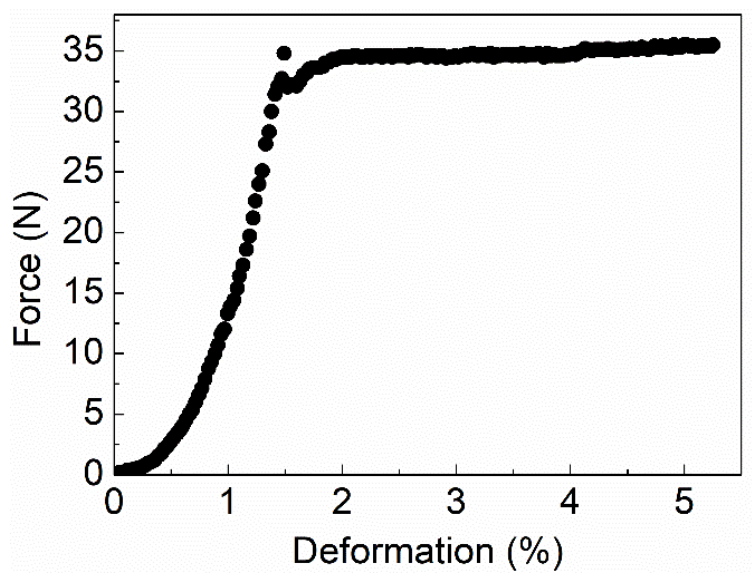

**Figure S2.** Force-deformation plot of a CDD substrate measured at 26°C under compression. Compression rate: 0.5 mm/min.

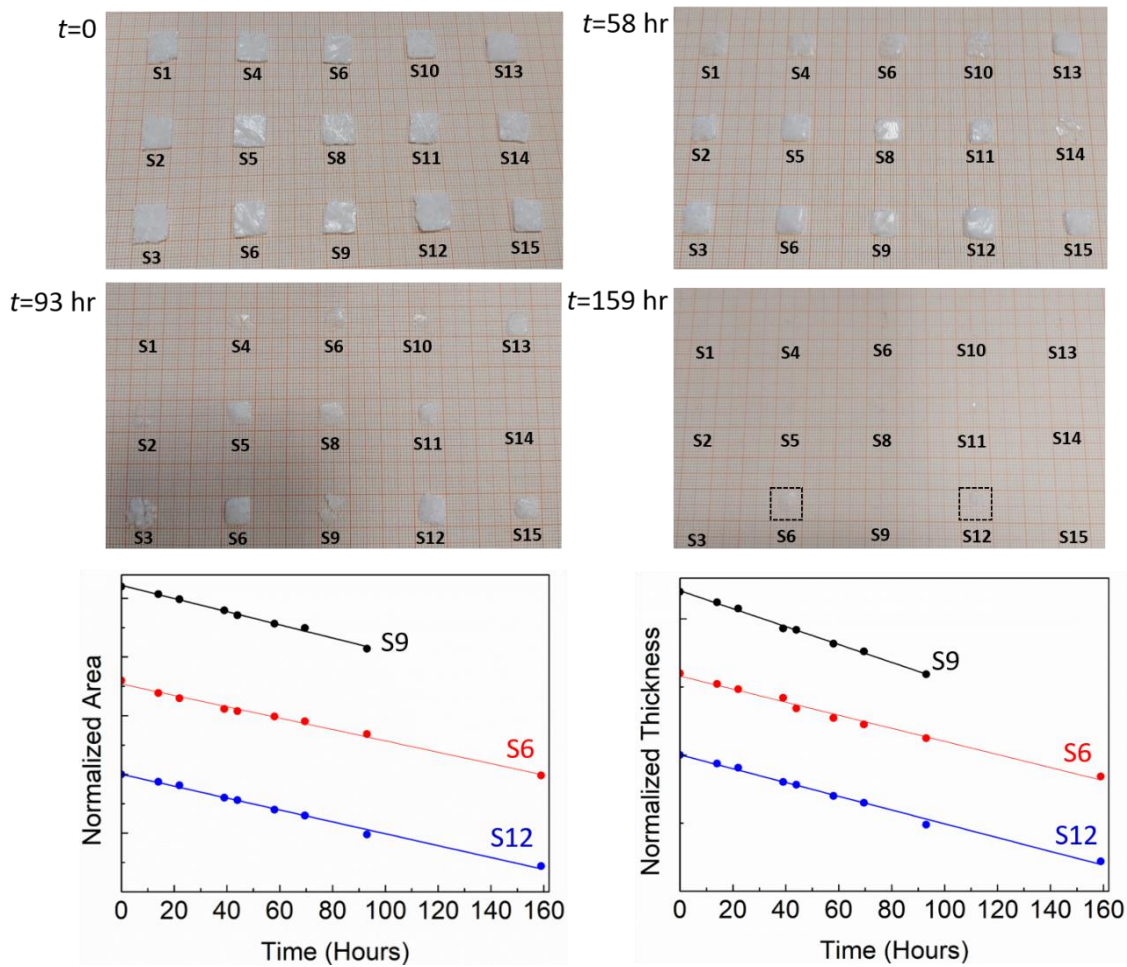

**Figure S3.** Top photographs: CDD substrates (S1-S15) captured at different instant times (*t*=0, 58, 93, 159 hours) during sublimation at room temperature under a fume hood with a face velocity of 1.9 m s$^{-1}$. Bottom plots show the normalized area (left) and thickness (right) *vs.* sublimation time, for the samples labelled by S6, S9 and S12 (data vertically-shifted for better clarity).

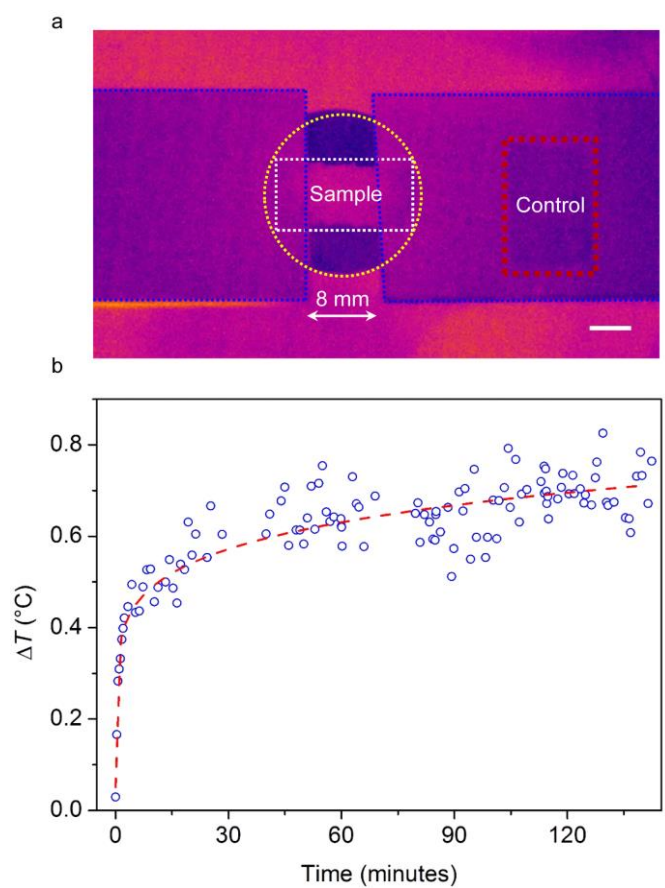

**Figure S4**. (a) Infrared thermal image of a CDD sample (white box), illuminated under an optical microscope (illuminated region: yellow dashed circle). The sample is suspended between two glass slides, highlighted by blue dashed lines. A second CDD sample (red box), positioned in a region that is not illuminated, serves as reference for determining the illumination-induced temperature variation ($\Delta T$). This thermal photograph is collected after 15 minutes of continuous irradiation. Scale bar: 5 mm. (b) Temporal evolution of $\Delta T$ during illumination (circles), evidencing an overall temperature increase below 1°C. The dotted line is a guide for the eye.

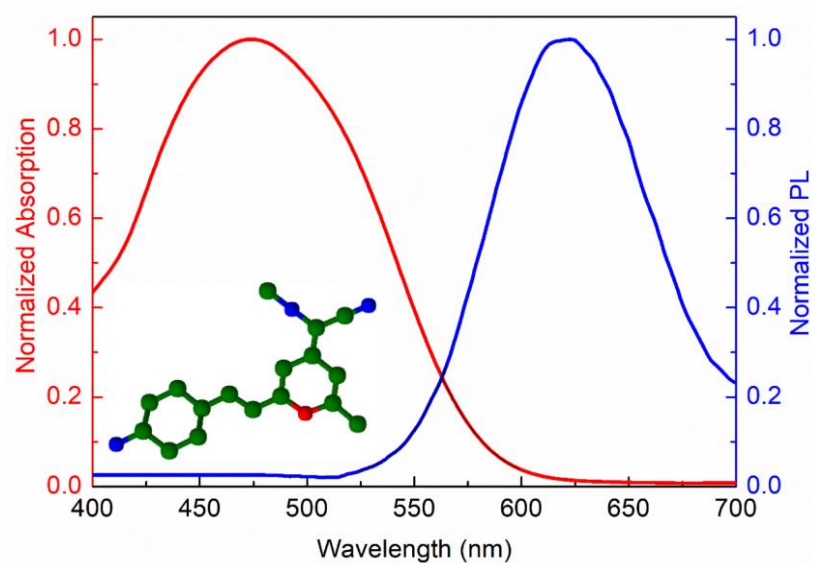

**Figure S5.** Absorption (red line, left vertical scale) and PL (blue line, right vertical scale) spectra of a PVA/DCM:PVP bilayer deposited on a quartz substrate. Dye-doping in PVP is carried out at 1% wt:wt DCM:PVP concentration. Inset: chemical structure of DCM (Colours indicate different atoms. Red: oxygen, green: carbon, blue: nitrogen).

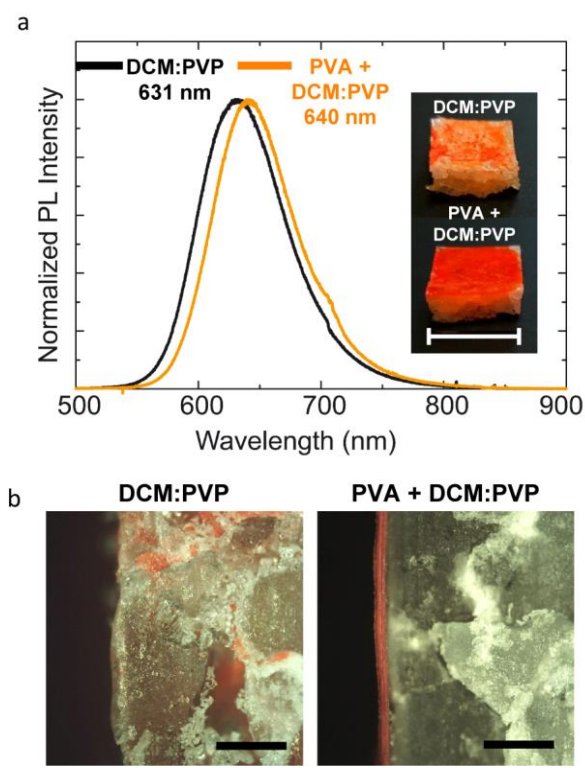

**Figure S6.** Comparison between films of PVP doped with 1% wt:wt of DCM (DCM:PVP) deposited either directly on top of CDD or on an intermediate PVA layer. (a) PL and photographs of the samples highlighting a blue-shift of the emission and a permeation of DCM:PVP into the CDD in absence of the PVA. Scale bar: 1 cm (b) Optical micrographs of the sample cross sections, supporting DCM:PVP permeation in the substrate without PVA, and showing the planarizing effect of the PVA coating. Scale bar: 400 µm.

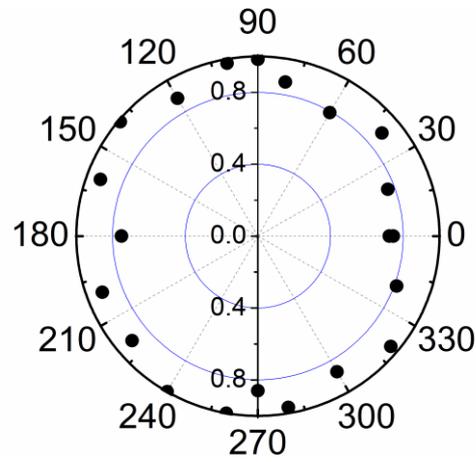

**Figure S7.** Polar plot of the normalized ASE intensity (circles) as a function of the angle of the analyzer polarization filter axis, $\theta$, measured with respect to the device thickness axis ($\theta = 0°$ for polarizer axis parallel to the device thickness axis, $\theta = 90°$ for polarizer axis perpendicular to the device plane).

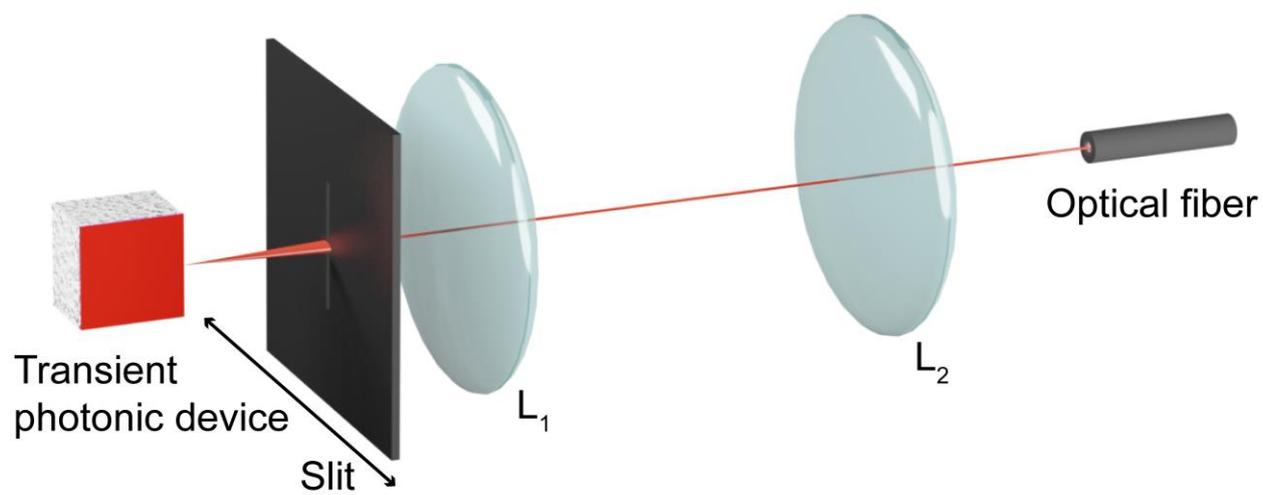

**Figure S8.** Experimental geometry used for measuring the ASE beam divergence.

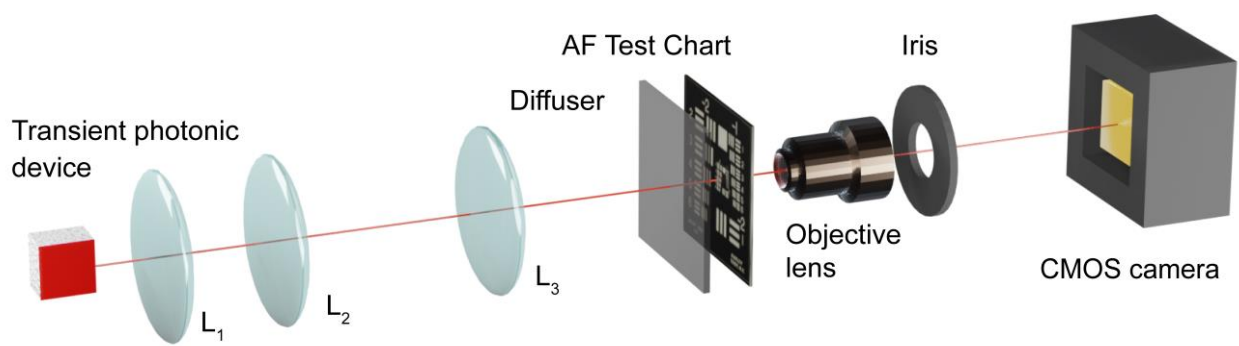

**Figure S9.** Scheme of the experimental set-up used for full-field imaging experiments.

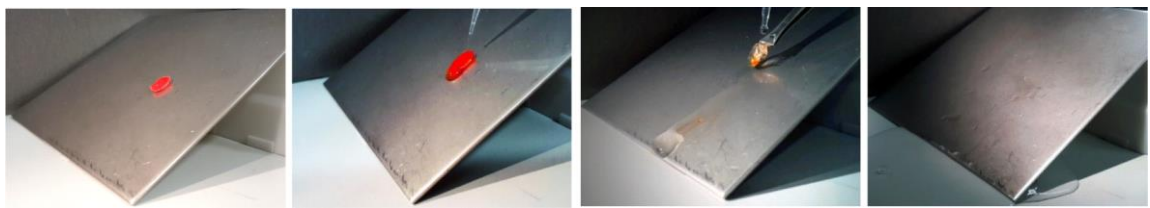

**Figure S10.** Photographs of dissolution of the residual polymer layers in water, at representative instants.